\documentclass[aps,twocolumn,superscriptaddress,10pt,floatfix,prl]{revtex4-1}

\usepackage{amssymb,amsmath}
\usepackage{graphicx}
\usepackage[colorlinks,linkcolor=blue,citecolor=blue,urlcolor=blue]{hyperref}
\usepackage{pdfpages} % include appendix
\makeatletter
\AtBeginDocument{\let\LS@rot\@undefined}
\makeatother

\begin{document}

\title{Commensurate-Incommensurate Transitions of the 1D Disordered Chiral Clock Model}

\author{Pengfei Liang}
\affiliation{Beijing Computational Science Research Center, Beijing 100193, China}

\author{Rosario Fazio}
\affiliation{The Abdus Salam International Center for Theoretical Physics, Strada Costiera 11, 34151 Trieste, Italy}
\affiliation{Dipartimento di Fisica, Universit{\`a} di Napoli ``Federico II'', Monte S. Angelo, I-80126 Napoli, Italy}
\affiliation{Beijing Computational Science Research Center, Beijing 100193, China}

\author{Stefano Chesi}
 \email{stefano.chesi@csrc.ac.cn}
\affiliation{Beijing Computational Science Research Center, Beijing 100193, China}
\affiliation{Department of Physics, Beijing Normal University, Beijing 100875, China}

\begin{abstract}
We study the effects of quenched disorder on the commensurate-incommensurate transitions in the 
1D $\mathbb{Z}_N$ chiral clock model. The interplay of domain walls and rare regions rounds 
the sharp transitions of the pure model. The density of domain walls displays an essential singularity, 
while the order parameter develops a discontinuity at the transition. We perform extensive 
density-matrix renormalization group calculations to support theoretical predictions. Our results 
provide a distinct rounding mechanism of continuous phase transitions in disordered systems.
\end{abstract}

\date{\today}

\maketitle

Understanding phase transitions in disordered systems is a long-standing 
problem in condensed matter physics. Quenched disorder that only causes fluctuations of the 
local critical point without explicitly breaking the underlying symmetries 
is of random-mass type \cite{Vojta_2006,Vojta2019}. 
The Harris criterion \cite{Harris_1974} states that weak random-mass disorder 
is irrelavent to a clean critical point if $\nu\ge2/d$, where $\nu$ is the correlation 
length exponent and $d$ is the spatial dimensionality. If disorder is relevant, the 
clean critical point flows to a new dirty fixed point characterized 
by either finite \cite{PhysRevLett.89.177202,PhysRevLett.93.097201,
PhysRevB.74.094415,PhysRevB.94.134501,PhysRevB.85.094202,PhysRevB.94.134501} 
or infinite \cite{McCoyWu1968prl,Fisher1992prl,Fisher1995prl} randomness.  
Another crucial aspect of disordered phase transitions is the role
played by rare regions, such as large spatial regions which are
locally ordered. Recently, a classification of critical points has been proposed \cite{Vojta_2006,Vojta2019}, depending on the effective dimensionality $d_r$ of rare regions and the lower critical dimension $d_c^-$ of the phase transition. If $d_r\le d_c^-$, a (quantum) Griffiths region is formed on both sides of the dirty critical point, where some physical quantities display essential or power-law Griffiths singularities \cite{Griffiths1969prl}; 
while if $d_r>d_c^-$, local static order can develop in rare regions, rounding 
the original sharp transition \cite{PhysRevLett.90.107202,PhysRevB.69.174410}.
Disorder can also round classical or quantum first-order transitions 
to continuous ones~\cite{PhysRevLett.62.2503,PhysRevLett.100.015703}. 
In the context of 1D $N$-state clock models, the strong disorder renormalization group (SDRG) method has been applied to the \emph{achiral} case \cite{IGLOI2005277}, showing that the critical point flows either to infinite randomness fixed points for $N=3,4$, where disorder dominates over quantum fluctuations at large distance, or to clean fixed points for $N\ge5$, where weak disorder is irrelevant. This is consistent with the Harris criterion since for $N=3,4$ phase transitions in the corresponding clean model are in the $3$-state Potts and Ashkin-Teller universality classes respectively, both of which satisfy $\nu<2$. 

On the other hand, to the best of our knowledge, disorder in the chiral clock model (CCM), a variant breaking charge conjugation symmetry, remain unexplored. Interest in this model comes from the fact that a new commensurate-incommensurate (C-IC) transition occurs \cite{schulz1980prb,ostlund1981prb,zhuangprb2015,sachdevpra2018,sachdevprb2018}, which is known to violate the Harris criterion. More specifically, the transition is of the Pokrosky-Talapov (PT) type \cite{PT1979prl} with $\nu=1/2$ for $N\ge4$, and always happens through an intermediate incommensurate (IC) critical phase. For $N=3$, both numerical simulations and field-theoretical analysis support a direct transition with a continuous varying exponent $\nu<5/6$ between ordered and disordered phases in the vicinity of the Potts point \cite{zhuangprb2015,sachdevpra2018,sachdevprb2018}, which then crosses over to the PT transition. The magnetic ordered phase is a domain-wall-free commensurate (C) phase and there might be both clockwise (CC) and counterclockwise (CW) domain wall excitations on top of it near the C-IC transition. Breaking of charge conjugation makes one species of domain walls energetically more favored, thereby an excess amount of domain walls destroys the magnetic long-range order (LRO) in the IC phase.

\begin{figure}
\begin{center}
\includegraphics[clip,width=0.48\textwidth]{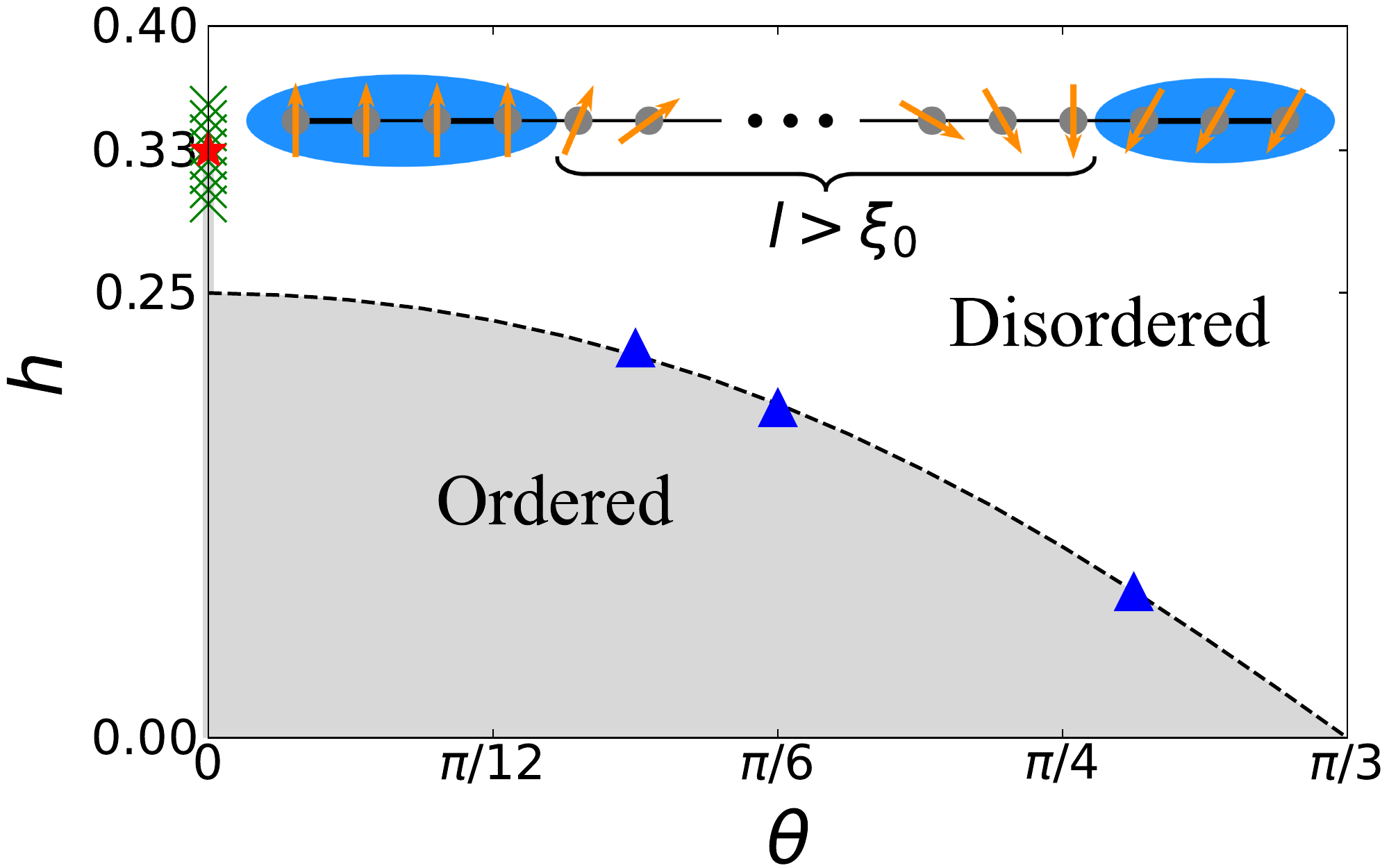}%
\end{center}
\caption{Phase diagram of the $N=3$ model, with $J_w=1/4$, $J_s=1$, and $\phi=0$. The dashed line is the phase boundary $h_c^w$ of the clean model without strong links, i.e., $p=0$ in Eq.~(\ref{distribution}). Red and blue symbols are critical points with finite disorder ($p=0.2$), see the dashed lines in Fig.~\ref{fig:crossings}(c). The green crosses at $\theta=0$ indicate the presence of a Griffith region around the critical point. Inset: for $h>h_c^w$ rare regions of weak links support a finite density of domain walls, destroying long-range order. $\xi_0$ is the correlation length at $p=0$.}
\label{fig:phase_diagram}
\end{figure}

In this work, we aim to study the effects of random-mass disorder on the C-IC transitions 
in the 1D $\mathbb{Z}_N$ CCM by means of various theoretical arguments 
and density-matrix renormalization group (DMRG) simulations. 
Our main findings can be summarized as follows: excess domain-wall excitations localized in rare regions of weakly-coupled sites (illustrated in the inset of Fig.~\ref{fig:phase_diagram}) round the sharp transition and, interestingly, induce a nonuniversal essential singularity in the density of domain walls, while the $\mathbb{Z}_N$ order parameter develops a discontinuity at 
the transition. The latter does not mean that the transition is first-order as there is no 
phase coexistence. Our results are in sharp contrast with the corresponding achiral model 
and show a different rounding mechanism compared to Refs.~\cite{PhysRevLett.90.107202,PhysRevB.69.174410}.

The $\mathbb{Z}_N$ random chiral clock model in 1D is given by the Hamiltonian,
\begin{align}\label{eq:hamiltonian}
H(\theta,\phi)=-\sum_{j} \left(J_j \sigma_j^\dagger \sigma_{j+1} e^{i\theta} + h_j \tau_j^\dagger e^{i\phi} + {\rm H. c.} \right),
\end{align}
defined on a chain of length $L$. The bonds $J_j$ and transverse fields $h_j$ are taken to be independent identically distributed (i.i.d.) random variables and we assume they are positive, $J_j,h_j>0$. The $N$-state spin operators $\tau$ and $\sigma$ on each site obey the algebra $\tau^N=\sigma^N=1, \; \sigma\tau=\omega \tau\sigma, \; \tau^\dagger = \tau^{-1}, \; \sigma^\dagger = \sigma^{-1}$,
with $\omega=e^{i2\pi/N}$. Unlike the Pauli operators for a spin-half, these operators are non-Hermitian with complex eigenvalues 
$\omega^0, \omega^1, \cdots, \omega^{N-1}$. In the $\sigma$-representation, $\sigma$ indicates the direction of spin while $\tau$ rotates the spin clockwise by a discrete angle $2\pi/N$. A nonzero $\theta$ or $\phi$ breaks the charge conjugation symmetry~\footnote{See Supplemental Material for discussion of the symmetries, the C-IC transition and its low-energy effective Hamiltonian, the decimation rules as well as for more DMRG results}, which turns out to be vital in the description of the transitions. The clean model possesses a $\mathbb{Z}_N$ symmetry-broken phase with finite magnetization, whose phase boundary for $N=3$ and $\phi=0$ is shown in Fig.~\ref{fig:phase_diagram}.

%%%%%%%%%%%%%%%%%%%%%%%%%%%%%%%%%%%%%%%%%%%%%%%%%%%%%
%%%%%%%%%%%%%%%%%%%%%%%%%%%%%%%%%%%%%%%%%%%%%%%%%%%%%

%%%%%%%%%%%%%%%%%%%%%%%%%%%%%%%%%%%%%%%%%%%%%%%%%%%%%
\begin{figure}%[!htbp]
\begin{center}
\includegraphics[clip,width=0.48\textwidth]{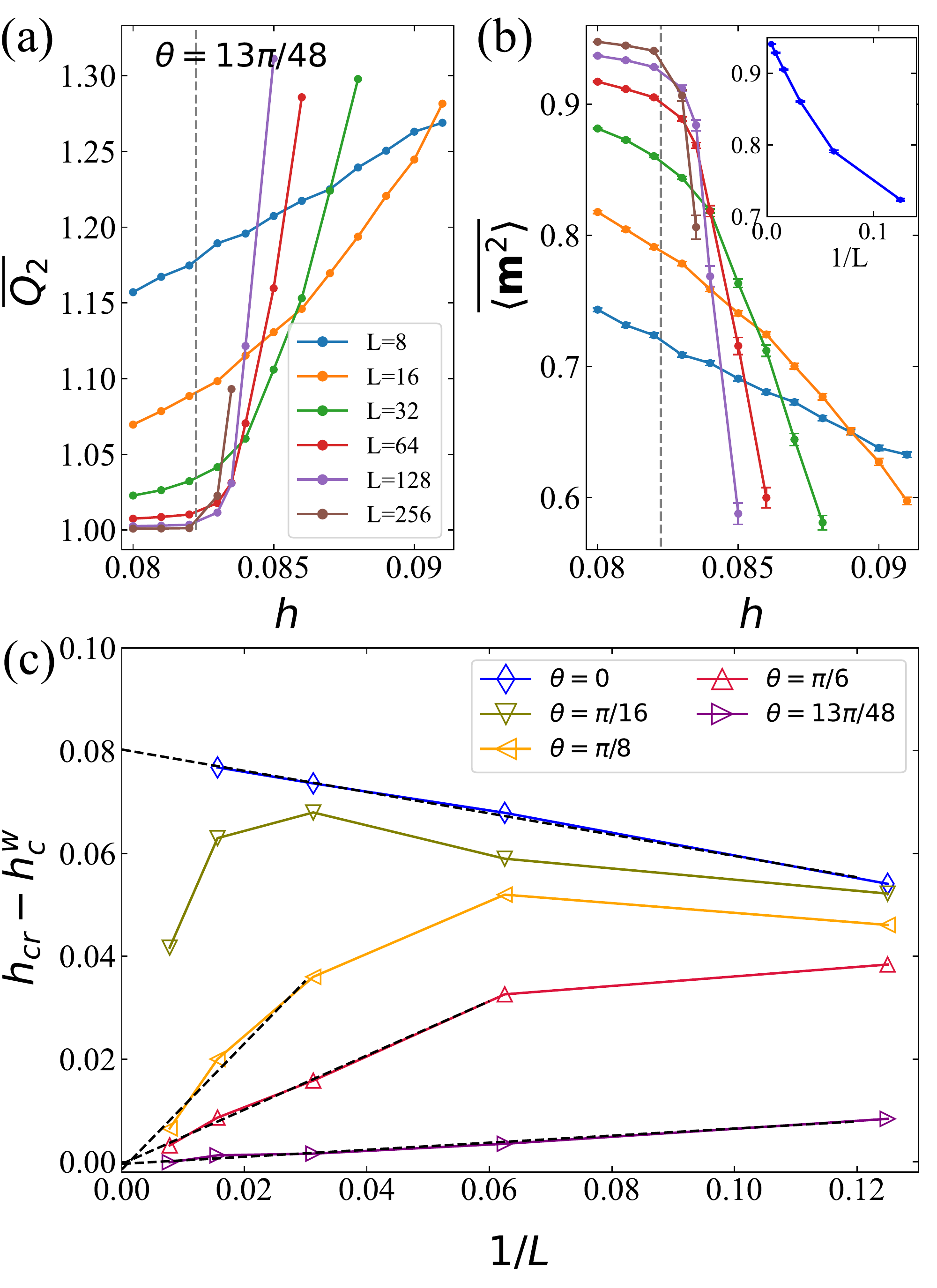} 
\end{center}
\caption{(a) $\overline{Q_2}$ and (b) $\overline{\langle\protect\mathbf{m}^2\rangle}$ as a function of $h$ at $\theta=13\pi/48$ and different values of $L$. The dashed lines mark $h_c^w$. Inset of panel (b): $\overline{\langle\protect\mathbf{m}^2\rangle}$ plotted against $1/L$ at $h=h_c^w$. Panel (c) shows the crossing points $h_{cr}$ between $\overline{Q_2}$ curves of neighboring system size ($L$ and $2L$), at several values of $L$ and $\theta$. For clarity, the clean critical value $h_c^w$ has been subtracted from $h_{cr}$. In all panels we take $J_s=1$, $J_w=1/4$, and  $p=0.2$.}
\label{fig:crossings}
\end{figure}
%%%%%%%%%%%%%%%%%%%%%%%%%%%%%%%%%%%%%%%%%%%%%%%%%%%%%

In the rest of the paper we assume $\phi=0$ and, by symmetry considerations~\cite{Note1}, restrict the range of $\theta$ from $0$ to $\pi/N$. We also consider uniform $h_j=h$ and, initially, the binary probability distribution: 
\begin{equation}\label{distribution}
P(J)=p\delta(J-J_s)+(1-p)\delta(J-J_w),
\end{equation}
where $J_{s(w)}$ indicates strong (weak) bonds and $p$ is the concentration of strong bonds. In analyzing the effects of disorder through DMRG~\cite{itensor}, we have chosen a rather small truncation parameter $\sim10^{-11}$ to ensure convergence in all calculations, and average over a sufficient number of disorder realizations (up to $10,000$ in selected cases), to achieve small statistical error bars. We address numerically the C-IC transition through the disorder-averaged $\overline{\langle\mathbf  m^2\rangle}$, with $\mathbf m=\sum_j \mathbf m_j /L$ and $\mathbf m_j = ((\sigma_j+\sigma_j^\dagger)/2, (\sigma_j-\sigma_j^\dagger)/2i)^T$, 
as well as the Binder ratio $\overline{Q_2}=\overline{\langle\mathbf  m^4\rangle}/\overline{\langle\mathbf  m^2\rangle}^2$,
where disorder average is done first to correctly detect phase transitions in disordered systems~\cite{lin2017prb}. Figure~\ref{fig:crossings} shows values of $\overline{\langle\mathbf  m^2\rangle}$ and $\overline{Q_2}$ computed at $N=3$ and $\theta=13\pi/48$. As expected, the system remains ordered for $h<h_c^w$, i.e., below the clean critical point associated with the weak links. In this region, the magnetization extrapolates to a finite value (see inset). On the other hand, as we will see shortly, the numerical data support the occurrence of a disordered phase with the same phase boundary $h_c^w$ of the clean model. 

To obtain the disordered critical point, we analyze the $\overline{Q_2}$ curves of Fig.~\ref{fig:crossings}(a), whose crossing points shift towards $h_c^w$ at large $L$. As shown in Fig.~\ref{fig:crossings}(c), we have examined other values of $\theta$ and observed a similar behavior: By plotting the difference between crossing points $h_{cr}$ and $h_c^w$ as function of inverse length $1/L$, we find that in all cases (if $\theta>0$) $h_{cr}$ shifts towards $h_c^w$ beyond a certain crossover length scale $L^*$. Remarkably, for $\theta \geq \pi/8$ extrapolation of data with $L>L^*$ (dashed lines) gives values in close agreement with $h_c^w$, also shown by the blue triangles in Fig.~\ref{fig:phase_diagram}. For $\theta= \pi/16$, the large size of the crossover scale $L^*$ does not allow for an accurate extrapolation, but the behavior in Fig.~\ref{fig:crossings}(c) is still consistent with the $h_c^w$ phase boundary.

%%%%%%%%%%%%%%%%%%%%%%%%%%%%%%%%%%%%%%%%%%%%%%%%%%%%%
\begin{figure}%[!htbp]
\begin{center}
\includegraphics[clip,width=0.48\textwidth]{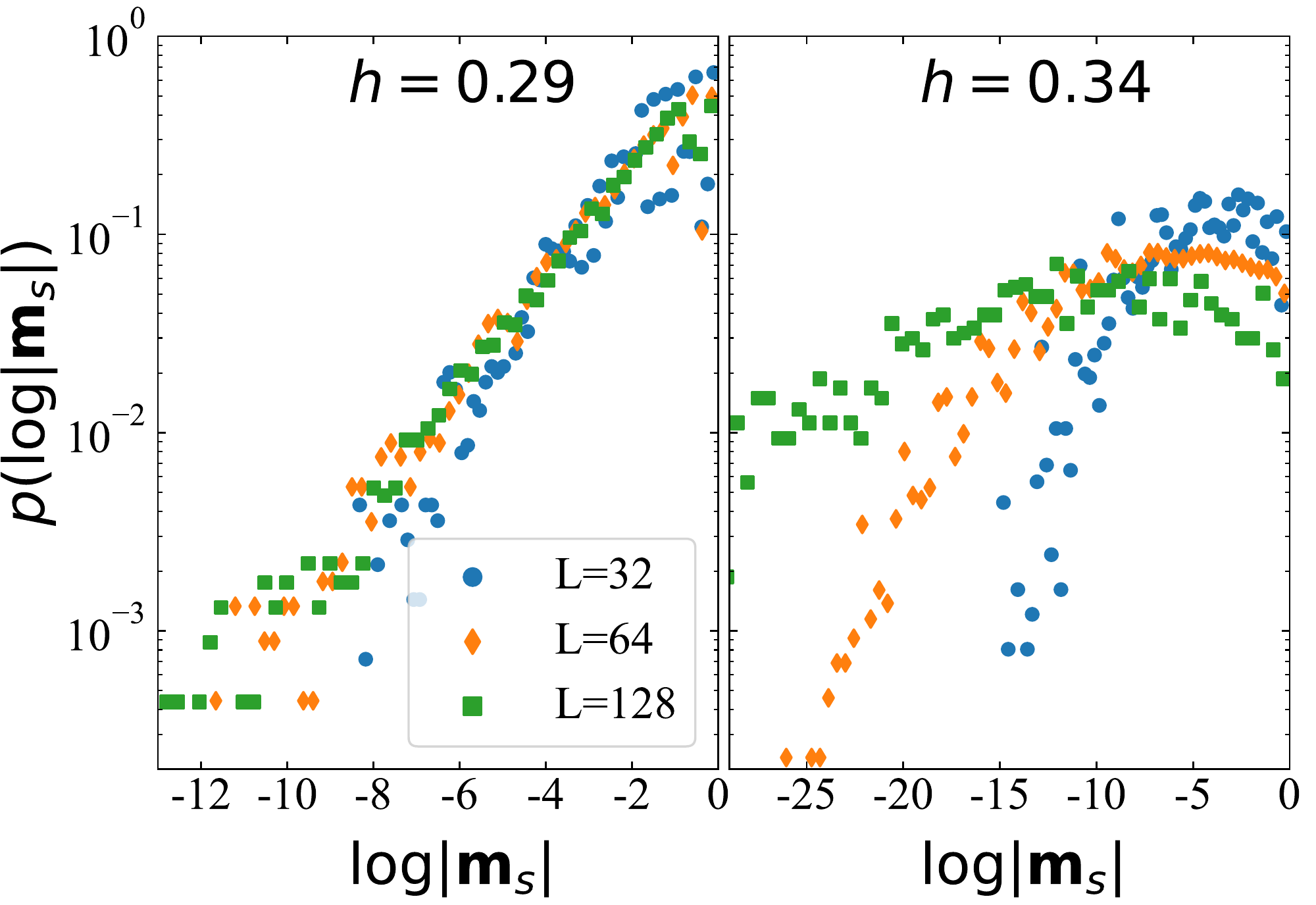} 
\end{center}
\caption{Distribution of the surface magnetization ${\bf m}_s$ at two different values of $h$. Other parametrs are as in Fig.~\ref{fig:crossings}.}
\label{fig:surface_magnetization}
\end{figure}
%%%%%%%%%%%%%%%%%%%%%%%%%%%%%%%%%%%%%%%%%%%%%%%%%%%%%

For $\theta=0$, the monotonic behavior of $h_{cr}$ extrapolates to $h_c^* \approx 0.33$, close to the theoretical predication for the random transverse field Ising chain (TFIC), $h_c^{\rm TFIC}=\exp[p\log J_s+(1-p)\log J_w]$ \cite{PhysRevLett.86.1343}. As the latter is controlled by an infinite randomness fixed point (IRFP), we test if the same is true in our model. Under this scenario, an independent assessment of $h_c^*$ is through the surface magnetization, computed by fixing the state of spin $L$ to one of the eigenstates of $\sigma_L$ (say, $|0\rangle$) and defining $\mathbf m_s=\mathbf m_1$. For an IRFP, the distribution $p(\log|\mathbf  m_s|)$ features an exponential tail in the ordered Griffiths region~\cite{PhysRevLett.86.1343}:
\begin{equation}
p(\log|\mathbf  m_s|) \sim e^{1/z \log|\mathbf  m_s|}, 
\end{equation}
where $z$ is the dynamical exponent and satisfies $z>1$. As shown in Fig.~\ref{fig:surface_magnetization}, convergence to an exponential scaling is indeed achieved at $L=128$ for values of $h$ as large as $h\approx 0.29$, while broadening and drifting of $p(\log|\mathbf  m_s|) $ towards smaller magnetization are found when $h> h_c^{\rm TFIC}$. Such behavior provides clear evidence of an IRFP with $h_c^*> 0.29$ but a precise estimate of the critical point through $p(\log|\mathbf  m_s|) $ is challenging. In \cite{Note1} we perform a detailed comparison of the CCM and TFIC, showing that much larger values of $L$ (only available in the TFIC) are necessary to obtain exponential scaling when $h \approx h_c^{\rm TFIC}$. On the other hand, the close similarity of $p(\log|\mathbf  m_s|)$ between the two models at accessible system sizes \cite{Note1}, together with the finite-size scaling analysis of Fig.~\ref{fig:crossings}(c), support the conjecture $h_c^*= h_c^{\rm TFIC}$. 

In the phase diagram of Fig.~\ref{fig:phase_diagram}, the value $h_c^*\approx h_c^{\rm TFIC}$ is marked as a red star. Physically, the singular behavior at $\theta=0$ is allowed by the divergence of the crossover scale $L^*$ at vanishing $\theta$. An argument based on the flow of the SDRG, indicating that the limits $L\to \infty$ and $\theta \to 0$ do not commute, will be discussed towards the end.

%%%%%%%%%%%%%%%%%%%%%%%%%%%%%%%%%%%%%%%%%%%%%%%%%%%%%
\begin{figure}%[!htbp]
\begin{center}
\includegraphics[clip,width=0.48\textwidth]{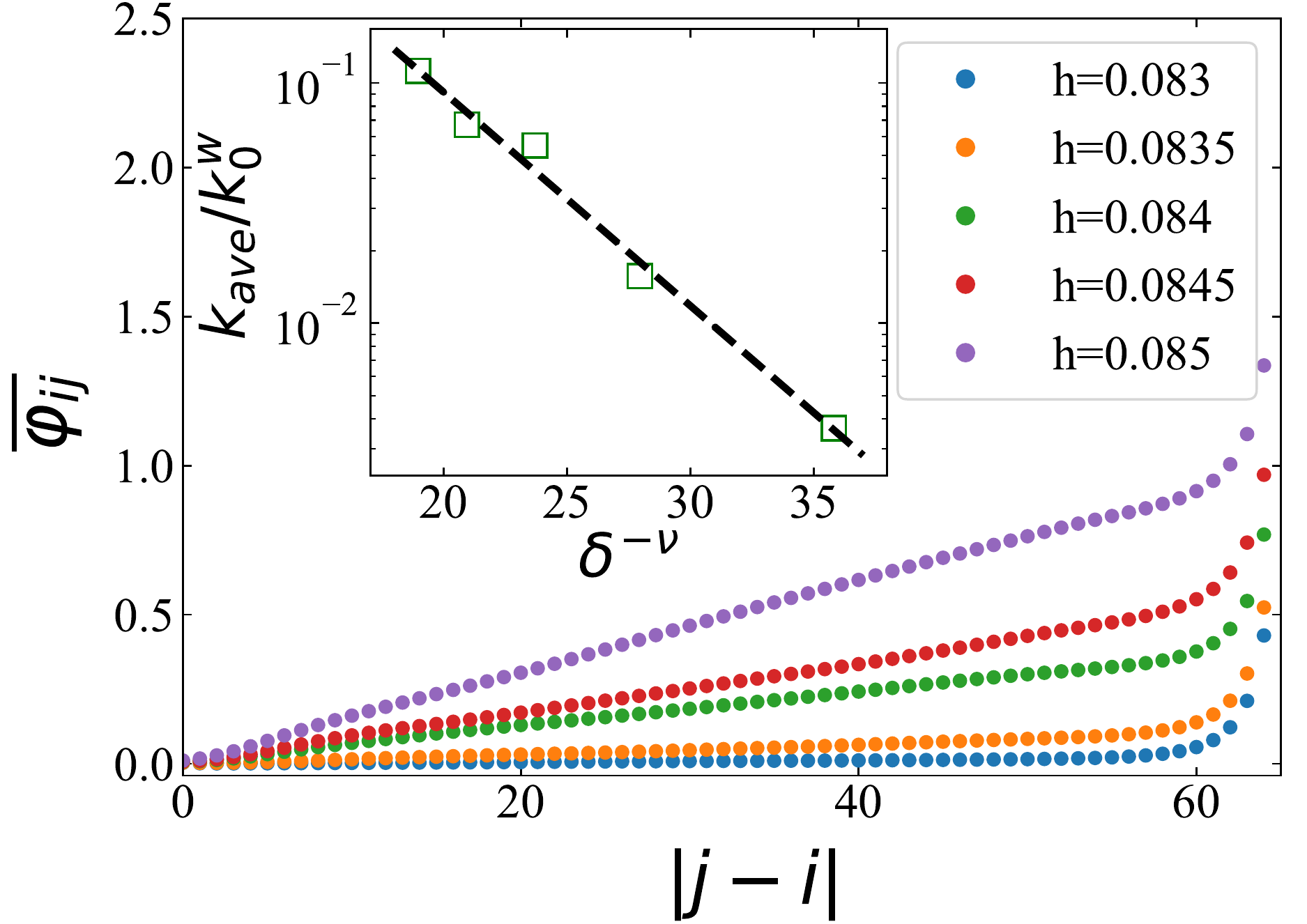} 
\end{center}
\caption{
Average angle $\overline{\varphi_{ij}}$ vs distance $|j-i|$ for different values of $h$. The inset shows the scaling of $k_{ave}/k_0^w$ with $\xi_0 \propto \delta^{-\nu}$, where $k_0^w$ is obtained from simulations of the clean model (see Fig.~\ref{fig:phase_transition}) and $\nu$ is from Ref.~\cite{sachdevpra2018}. We used $J_s=1$, $J_w=1/4$, $p=0.2$, $L=128$, and $\theta=13\pi/48$.}
\label{fig:crossing_singularity} 
\end{figure}
%%%%%%%%%%%%%%%%%%%%%%%%%%%%%%%%%%%%%%%%%%%%%%%%%%%%% 

We now return to the chiral case, recalling that at $\theta\neq 0$ the phase transition features two complementary order parameters: magnetization and domain wall density $\rho$. In the clean model, a finite amount of excess CC (or CW) domain walls will completely destroy the magnetic order, and the transitions of $\langle\mathbf m^2\rangle$ and $\rho$ happen simultaneously. Domain walls cause winding of spins in space, thus $\rho$ is accessible through the angle $\varphi_{ij}$ between two spins $i,j$. By using $\cos\varphi_{ij} \propto \langle \mathbf m_i\cdot\mathbf m_j\rangle = \mathrm{Re}\langle \sigma_i \sigma_j^\dagger \rangle$, where the expectation is taken over a $\mathbb{Z}_N$ symmetric state, we obtain $\varphi_{ij}$ by calculating the two-points correlator $\langle \sigma_i \sigma_j^\dagger \rangle = A_{ij} e^{i\varphi{ij}}$, where $A_{ij}$ is the amplitude. In the clean case, a linear behavior of type $\varphi_{ij}\simeq k_0 (i-j)$ is obtained in the bulk of the chain, where the slope $k_0$ is a measure of the density of excess domain walls. It vanishes in the absence of domain walls (as in the ordered phase) or if the two species of domain walls are excited equally (at $\theta=0$, guaranteed by the charge conjugation symmetry). As shown in Fig.~\ref{fig:crossing_singularity}, in the disordered model we find a similar linear dependence for the average angle $\overline{\varphi_{ij}}\simeq k_{ave}(i-j)$.

Consistent with previous findings, $k_{ave}$ approaches zero at the clean phase boundary $h_c^w$, where the leading singularity may be inferred through Lifshitz-tail arguments. As illustrated in Fig.~\ref{fig:phase_diagram} (inset), domain walls are most likely to live in extended regions of length $l\gg 1 $ composed of only weak links. Since $J_j$ are independent random variables, the probability that a site belongs to one of such regions follows the Poisson distribution $f(l) = e^{-\alpha l}(1-e^{-\alpha})$ with $\alpha=-\ln(1-p)$, and is exponentially suppressed with $l$. In these rare regions, the linear increase of $\overline{\varphi_{ij}}$ is controlled by $k_0^w$, i.e., the slope of the clean model with $J_j = J_w$. We thus estimate:
\begin{equation}\label{k_ave_estimate}
k_{ave} \simeq k_0^w\sum_{l=l_0}^\infty f(l) = k_0^w e^{-\alpha l_0}, 
\end{equation}
where $l_0$ is a suitable cutoff. It is natural to identify $l_0\sim c\xi_0$, with $\xi_0$ the clean correlation length. Indeed, by considering individual disorder realizations at $h>h_c^w$, we find that ordered domains are formed where strong links are dense. For a given $i$, the angle $\varphi_{ij}$ is basically pinned to a constant when $j$ belongs to such an ordered domain \cite{Note1}. Instead, as expected, magnetic correlations decay exponentially between two distant $J_s$ links, where $\varphi_{ij}$ follows approximately the linear dependence of the clean model. From these arguments and Eq.~(\ref{k_ave_estimate}), we conclude that $k_{ave}$ is given by:
\begin{equation}\label{eq:kave}
k_{ave}/k_0^w \sim e^{-\alpha c \xi_0}=e^{-\alpha c'\delta^{-\nu}},
\end{equation}
where $\delta = |h-h_0^w|$ measures the distance to $h_0^w$ and $\nu$ is the corresponding clean correlation length exponent \footnote{Note that Eq.~(\ref{eq:kave}) is consistent with the two clean limits: when $p\to0$ there are no strong links and $k_{ave}\to k_0^w$ while in the limit $p\to1$, the system is in the ordered phase at $h_{c}^w$ and $k_{ave}\to 0$.}.  In the inset of Fig.~\ref{fig:crossing_singularity} we show that the numerically evaluated ratio $k_{ave}/k_0^w$ scales exponentially with $\xi_0$, confirming the presence of an essential singularity. We have also checked Eq.~(\ref{eq:kave}) is insensitive to system size $L$ and holds for other values of $\theta$ \cite{Note1}. 

%%%%%%%%%%%%%%%%%%%%%%%%%%%%%%%%%%%%%%%%%%%%%%%%%%%%%
\begin{figure}%[!htbp]
\begin{center}
\includegraphics[clip,width=0.48\textwidth]{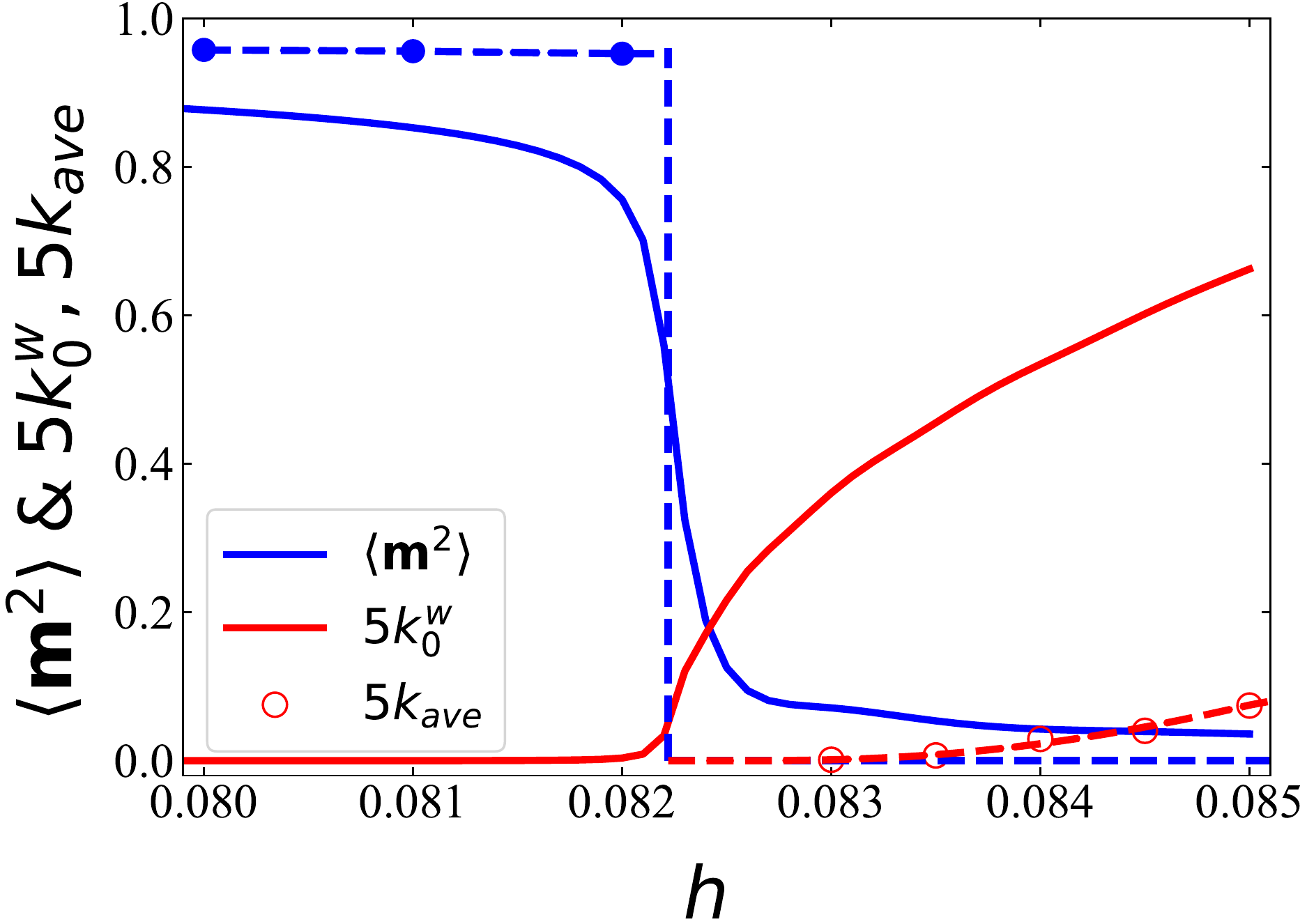} 
\end{center}
\caption{
Plot of $\langle \mathbf m^2\rangle$ and $k_0^w$ (solid lines) for the clean model with $J_j=J_w$. In the disordered case we show extrapolated values of $\overline{\langle \mathbf{m}^2 \rangle}$ (blue filled dots), obtained as in the inset of Fig~\ref{fig:crossings}(b), and $k_{ave}$ (red empty circles), obtained from Fig.~\ref{fig:crossing_singularity}. Dashed lines follow the theoretical predictions for the disordered case. All parameters are the same of Fig.~\ref{fig:crossing_singularity}.}
\label{fig:phase_transition} 
\end{figure}
%%%%%%%%%%%%%%%%%%%%%%%%%%%%%%%%%%%%%%%%%%%%%%%%%%%%% 

Figure~\ref{fig:phase_diagram} shows the remarkable difference in the critical properties of the two models (with and without disorder). Despite the jump in the averaged squared magnetization $\overline{\langle\mathbf  m^2\rangle}$, the domain wall density has a smooth dependence, thus the disordered phase transition is still continuous. In fact, as in a mixed-order phase transition~\cite{PhysRevLett.122.235701,PhysRevB.103.035108}, the correlation length $\xi$ of the diordered model diverges when approaching the critical point. By an argument similar to Eq.~(\ref{eq:kave}), we find an essential singularity also for $\xi$ \cite{Note1}. Below $h_c^w$ the presence of strong bonds implies that (as in Fig.~\ref{fig:phase_transition}) $\overline{\langle\mathbf  m^2\rangle }$ is larger than in the clean model. That $\overline{\langle\mathbf  m^2\rangle }>0$ at $h_c^w$ is a result of aligned locally ordered domains. For illustration, we can consider the simple limit of small $p$ (when strong links are rare). An isolated strong link nucleates a total magnetization $\Delta{\bf m}$ around it, thus a nonzero magnetization density $\sim p\Delta{\bf m}$ survives at $h \to h_c^w$. Importantly, when $h \lesssim h_c^w$ the intermediate regions formed by weak links support long-range order. Therefore, differently from the schematics in Fig.~\ref{fig:phase_diagram} (where $h>h_c^w$), two distant strongly ordered regions remain correlated.

The main features obtained using Eq.~(\ref{distribution}) are still valid for general $N$. Furthermore, the above arguments impose no constraint on the strength and functional form of disorder and apply to random $h_j$ as well. We have also analyzed Eq.~(\ref{eq:hamiltonian}) through a rigorous mapping of chiral domain-wall excitations to spinless fermions, valid for general $N$ and small $\delta=\pi/N - \theta$ \cite{Note1}. A main conclusion of that approach is that $J_j$ mimics a random potential with strength $\propto J_j$, thus excess domain walls are most likely localized in the rare regions while strong links act like energy barriers for the domain walls. This physical picture is consistent with the critical behavior discussed so far. We note, however, that the functional form of the singular part of $k_{ave}/k_0^w$ is nonuniversal and may depend on the details of the distribution. For distributions of $J_j$ with infinitesimal small bonds (e.~g., $J_w\to0$) we may expect that long range order is completely washed out, as rare regions exist for any finite $h$.  We have studied explicitly the uniform distribution:
\begin{equation}
P(J) = \left(\Theta(J)+\Theta(J_s-J)\right)/2J_s,
\end{equation}
with $\Theta(J)$ the Heaviside step function. Indeed, now the crossings in $\overline{Q_2}$ shift all the way down to zero \cite{Note1}.

Finally, it is interesting to look at the phase transition form the perspective of the SDRG method.  The decimation rules of the random $\mathbb{Z}_N$ CCM read \cite{Note1}:
\begin{align}
h_j' = \kappa_{\theta_j} \frac{h_jh_{j+1}}{J_j},  \quad \phi_j'= \phi_j+\phi_{j+1}, \label{eq:rules1} \\
J_{j-1}' = \kappa_{\phi_j} \frac{J_{j-1}J_{j}}{h_j}, \quad \theta_{j-1}' = \theta_{j-1}+\theta_{j},\label{eq:rules2}
\end{align}
where $2\kappa_{\alpha} = \sum_{s=\pm1}[\cos\alpha-\cos(\alpha+2s\pi/N)]^{-1}$ and we have introduced a position dependence of the chiral phases $\theta_j, \phi_j$, to incorporate the fact that they will change during the renormalization process. In the achiral case ($\theta_j=\phi_j=0$) we recover the constant prefactor $\kappa_0=1/[1-\cos(2\pi/N)]$, thus the SDRG flow proceeds as usual. For $N=3,4$, we have $\kappa_0<1$, meaning that the critical point flows to IRFP; while for $N\ge5$, we have $\kappa_0>1$, meaning that weak disorder is irrelevant \cite{IGLOI2005277}. If the initial values of $\theta,\phi$ are nonzero, however, the chiral phases evolve in a nontrivial manner since they accumulate over progressively larger domains through Eqs.~(\ref{eq:rules1}) and (\ref{eq:rules2}). Inevitably, resonances will occur at some renormalization step, indicating a breakdown of the renormalization group flow. For small starting values, $\theta,\phi \ll 1$, the initial flow follows closely the achiral case. Significant deviations and the appearance of resonances only occur above a certain scale which, we suggest, is related to the large crossover length $L^*$ observed in numerical simulations.

In conclusion, we have shown that the sharp C-IC transitions in the disordered $\mathbb{Z}_N$ CCM are rounded by the interplay of excess domain wall excitations and rare regions. 
Compared to the known rounding mechanism of continuous phase transitions in Ref.~\cite{PhysRevLett.90.107202}, our results reveal an important role played by low-energy excitations in certain disordered continuous phase transitions. 

\begin{acknowledgments}

P.L. acknowledges computational resources from the Beijing Computational Science Research Center. R.F. acknowledges partial financial support from the Google Quantum Research Award. R.F. research has been conducted within the framework of the Trieste Institute for Theoretical Quantum Technologies (TQT). S.C. acknowledges support from NSFC (Grants No. 11974040 and No. 12150610464), and NSAF (Grant No. U1930402). 

\end{acknowledgments}

\bibliography{RCCMrefs}

% Adding sup mat
\bigskip
\onecolumngrid
\newpage


\section*{Supplementary material}

# Commensurate-Incommensurate Transitions of the 1D Disordered Chiral Clock Model - Supplemental Material

Pengfei Liang

*Beijing Computational Science Research Center, Beijing 100193, China*

Rosario Fazio

*The Abdus Salam International Center for Theoretical Physics, Strada Costiera 11, 34151 Trieste, Italy*
*Dipartimento di Fisica, Università di Napoli "Federico II", Monte S. Angelo, I-80126 Napoli, Italy and*
*Beijing Computational Science Research Center, Beijing 100193, China*

Stefano Chesi[*]

*Beijing Computational Science Research Center, Beijing 100193, China and*
*Department of Physics, Beijing Normal University, Beijing 100875, China*

## I. SYMMETRIES OF THE MODEL

In this section, we review the symmetries of the $\mathbb{Z}_N$ CCM [1, 2]. The chiral phase plane $(\theta, \phi)$ of interest can be greatly reduced by symmetry considerations. First, the form of $H(\theta + \frac{2\pi}{N}n, \phi + \frac{2\pi}{N}m)$ is unchanged by a redefinition of spin operators

$$\tau_j' = \tau_j \omega^{-1}, \quad \sigma_j' = \sigma_j \omega^{(j \bmod N)} \tag{1}$$

which retains the algebraic relations of the new operators at the same time, thus implying that physical properties of $H(\theta + \frac{2\pi}{N}n, \phi + \frac{2\pi}{N}m)$ are the same as those of $H(\theta, \phi)$. Second, we note there are three extra discrete transformations which may retain the form of the Hamiltonian, namely charge conjugation $C$, parity $P$ and time reversal $T$. Charge conjugation is defined by its action on spin operators as

$$C\sigma_j C = \sigma_j^\dagger, \quad C\tau_j C = \tau_j^\dagger, \quad C^2 = 1, \tag{2}$$

which lead to the transformation $CH(\theta, \phi)C = H(-\theta, -\phi)$. Likewise time reversal is given by

$$T\sigma_j T = \sigma_j^\dagger, \quad T\tau_j T = \tau_j, \quad T^2 = 1, \tag{3}$$

which acts as $TH(\theta, \phi)T = H(\theta, -\phi)$. For clean model, there is an extra parity symmetry defined by the relations

$$P\sigma_j P = \sigma_{-j}, \quad P\tau_j P = \tau_{-j}, \quad P^2 = 1, \tag{4}$$

and its action on the Hamiltonian is $PH(\theta, \phi)P = H(-\theta, \phi)$, which however is explicitly broken in the presence of disorder. For the case $\phi = 0$ we are interested in, combining the $2\pi/N$ periodicity of $H(\theta, 0)$ in $\theta$ and charge conjugation symmetry $CH(\theta, 0)C = H(-\theta, 0)$, we can restrict the range of $\theta$ to the interval $[0, \pi/N]$. Lastly, the present model has a global $Z_N$ symmetry defined by the operator $G = \prod_j \tau_j$, which is spontaneously broken in the ordered phase.

## II. MORE DMRG RESULTS

*Achiral Case.*— Here we show numerical results at $\theta = 0$, supplementing evidences to support the conjecture $h_c^* = h_c^{\text{TFIC}}$ presented in the main text. In Fig. 1 we show the Binder ratio $\overline{Q_2}$ as a function $h$ for different sizes $L$ at $\theta = 0$. The crossings of $\overline{Q_2}$ are already shown in the main text as well as in the inset. Results of the surface magnetization $\mathbf{m}_s$ are given in Fig. 2, where we compare the distributions $p(\log |\mathbf{m}_s|)$ of TFIC (upper panels) with

---

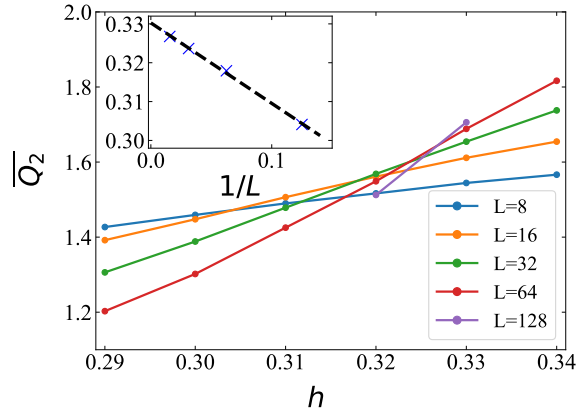

FIG. 1: $\overline{Q_2}$ as a function of $h$ for different $L$ in the achiral case. Inset shows the crossings (crosses) and a linear extrapolation (dashed line).

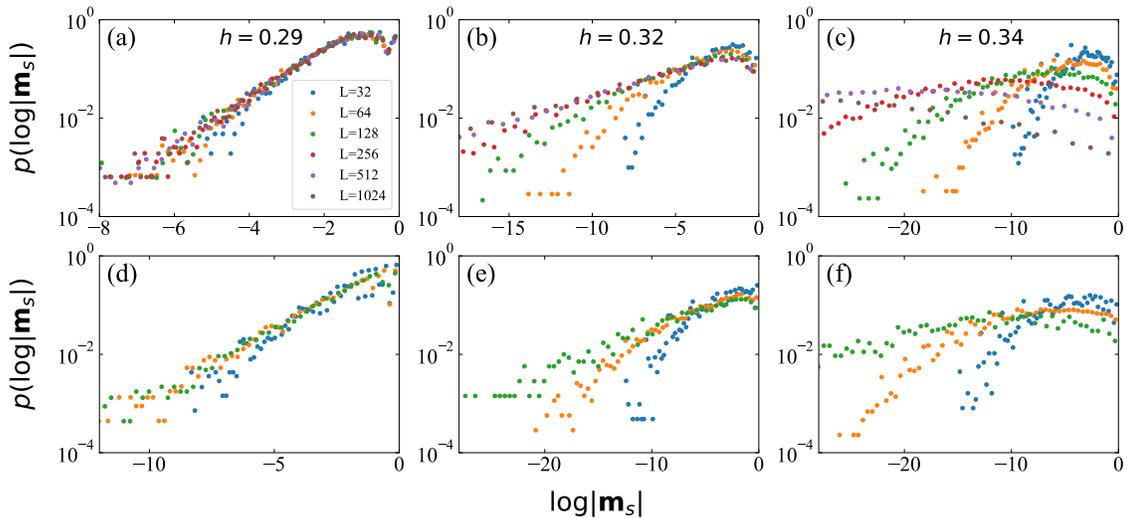

FIG. 2: Distributions $p(\log|\mathbf{m}_s|)$ of the surface magnetization for TFIC (upper panels) and CCM (lower panels) for various values of $h$.

that of CCM (lower panels) for h below (left and middle columns) and above (right column) the critical point. For TFIC, we employ the following theoretical formula [3]

$$|\mathbf{m}_s| = \left[1 + \sum_{l=1}^{L-1} \left(\frac{h_j}{J_j}\right)^2\right]^{-1/2},$$

which allows to reach much larger size. For $h = 0.29$, convergence is achieved at $L = 128$ in both models and as predicted $p(\log|\mathbf{m}_s|)$ features an exponential tail. For $h = 0.34$, broadening and drifting towards smaller magnetization of $p(\log|\mathbf{m}_s|)$ are clearly visible in both models. For $h = 0.32$, convergence for TFIC is only reached up to $L = 1024$, so it is hard to observe convergence for CCM using DMRG. However we can see the behaviors of $p(\log|\mathbf{m}_s|)$ are quite similar in both models. Therefore we are confident to say that these two independent numerical calculations coherently support the conclusion $h_c^* = h_c^{\mathrm{TFIC}} \approx 0.33$.

*Correlation Function in Individual Realization of Disorder.*— Fig. 3(a) shows a specific configuration of $J_j$ and the evaluated $A_{ij}$ and $\varphi_{ij}$ collected in our simluation for $\theta = \pi/8$ and $L = 128$. This plot (and many others not shown) supports the physical picture described in the main text: $\varphi_{ij}$ increases greatly in certain regions consisting of only weak bonds $J_w$ and is almost constant outside these regions. To examine how $\varphi_{ij}$ changes inside rare regions, we *artificially* vary the length $l$ of the gray region marked in Fig. 3(a) by inserting more weak bonds $J_w$ while traslating

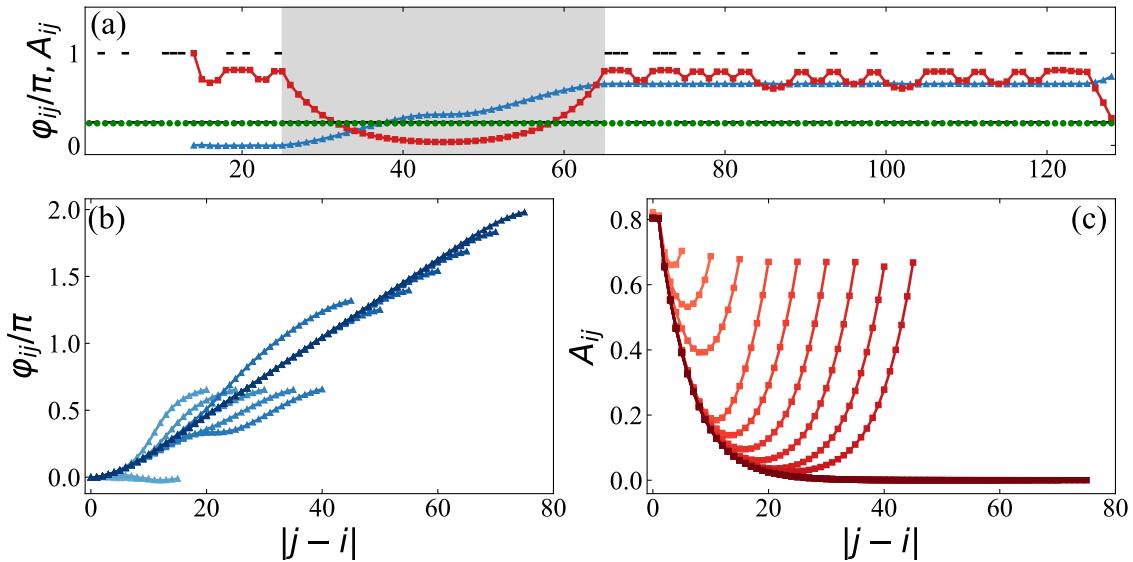

FIG. 3: (a) $\varphi_{ij}$ (blue) and $A_{ij}$ (red) for a specific realization of $J_j$. Here both $\varphi_{ij}$ and $A_{ij}$ are calculated from spin $i = 14$ for a chain of length $L = 128$. A rare region where $\varphi_{ij}$ increases greatly is marked by gray color. Green dots and black line segments indicate fields $h_j$ and bonds $J_j$ respectively. Other parameters: $\theta = \pi/8$, $J_s = 1$, $J_w = 1/4$, $p = 0.2$, $h = 0.24$. (b) and (c) show $\varphi_{ij}$ and $A_{ij}$ inside the gray region marked in panel (a) when the length of this region is varied, respectively.

the bonds on the right hand side. Note that we have chosen a rare region sandwiched by two large local domains where $\varphi_{ij}$ is supposed to be most strongly altered there. The results are shown in Fig. 3(b,c). As expected, there is a threshold length $l^*$ ($l^* \approx 18$ for the present realization) below which $\varphi_{ij}$ only displays small oscillations. Surprisingly, beyond $l^*$ we see that $\varphi_{ij}$ features a step-like function and its values at very step are $\varphi_{ij} = 2\pi/3n$ with $n = 1, 2, \cdots$. In the mean time, $A_{ij}$ displays completely revival. Only for very large rare regions, $\varphi_{ij}$ and $A_{ij}$ behave like the clean model. Nonetheless, our assumption of linear increasing of $\varphi_{ij}$ in rare regions used in the main text should still be valid as long as we are concerned only with the leading singularity of $k$. The complicated behavior observed here may cause certain subleading corrections to the scaling of $k$.

*Singularity of the Density of Excess Domain Walls.*— Fig. 4(a) shows $\overline{\varphi_{ij}}$ as a function of distance $|j - i|$ for several $h$ above $h_c^w \approx 0.0823$ computed at $\theta = 13\pi/48$ and $L = 64$. Note that to access long rare regions, we have computed $\overline{\varphi_{ij}}$ from spin $i = 10$. The scaling of $k_{ave}/k_0^w$ is plotted in the inset which also follows the exponential scaling relation. This confirms that the scaling form of $k_{ave}/k_0^w$ obtained in the main text is insensitive to system size $L$. In Fig. 4(b) we show $\overline{\varphi_{ij}}$ at $\theta = \pi/8$ for $h > h_c^w \approx 0.215$. Clearly the scaling form still holds which suggests the validity of our arguments for all nonzero $\theta$.

*$\overline{Q_2}$ and $\overline{\langle \mathbf{m}^2 \rangle}$ for other values of $\theta$.*— Fig. 5 shows more DRMG data of $\overline{Q_2}$ and $\overline{\langle \mathbf{m}^2 \rangle}$ for three other values of $\theta$. In all cases, we see the crossings of $\overline{Q_2}$ curves shift towards $h_c^w$ if $L$ exceeds $L^*$.

*Uniform Distribution.*— In Fig. 6 we show DMRG results of $\overline{Q_2}$ and $\overline{\langle \mathbf{m}^2 \rangle}$ for the case of uniform $h$ and random $J_j$ described by the probability distribution $p(J) = (\Theta(J) + \Theta(J_s - J))/2J_s$ with parameter $J_s = 1$ at $\theta = \pi/8$. Consistent with theoretical predication, the crossings of $\overline{Q_2}$ curves shift to zero, indicating that the transition point is $h_c = 0$.

*Singularity of $\xi$.*— The correlation function is $\langle \mathbf{m}_i \cdot \mathbf{m}_j \rangle = \text{Re} \langle \sigma_i \sigma_j^\dagger \rangle = A_{ij} \cos \varphi_{ij}$. As shown in Fig. 3(c), $A_{ij}$ crosses over to clean behavior in large rare regions. Consider a large number $n$ of weak-links regions with length $l_j \geq 1$, one after the other. Each of them is composed of $l_j - 1$ weak links and a strong link at the right boundary, therefore the probability of such regions is $f(l-1) = p(1-p)^{l-1}, l \geq 1$. After passing these regions, $A_r \equiv A_{|j-i|}$ falls as

$$A_r \sim \exp\left(-\sum_{l_j \geq l_0'} l_j/\xi_0\right) = \exp\left(-d/\xi_0\right), \tag{6}$$

where $d = \sum_{l_j \geq l_0'} l_j$ is the total length of rare regions with $l_0'$ a different cutoff, typically much larger than $l_0$ used for

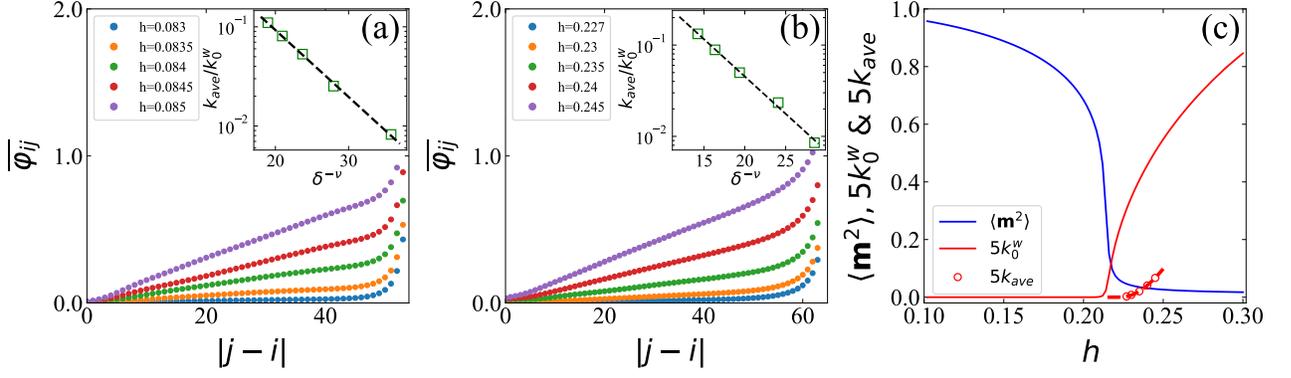

FIG. 4: $\overline{\varphi_{ij}}$ vs distance $|j-i|$ for different h at $\theta = 13\pi/48$ (a) and $\theta = \pi/8$ (b) calculated for $L = 64$ and $L = 128$ respectively. Insets shows the scaling of $k_{ave}/k_0^w$. Other parameters: $J_s = 1, J_w = 1/4, p = 0.2$. (c) $\langle \mathbf{m}^2 \rangle$ and $k_0^w$ of the clean model at $\theta = \pi/8$ and $L = 128$. Circles indicate $k_{ave}$ obtained by a linear fitting of the data in (b).

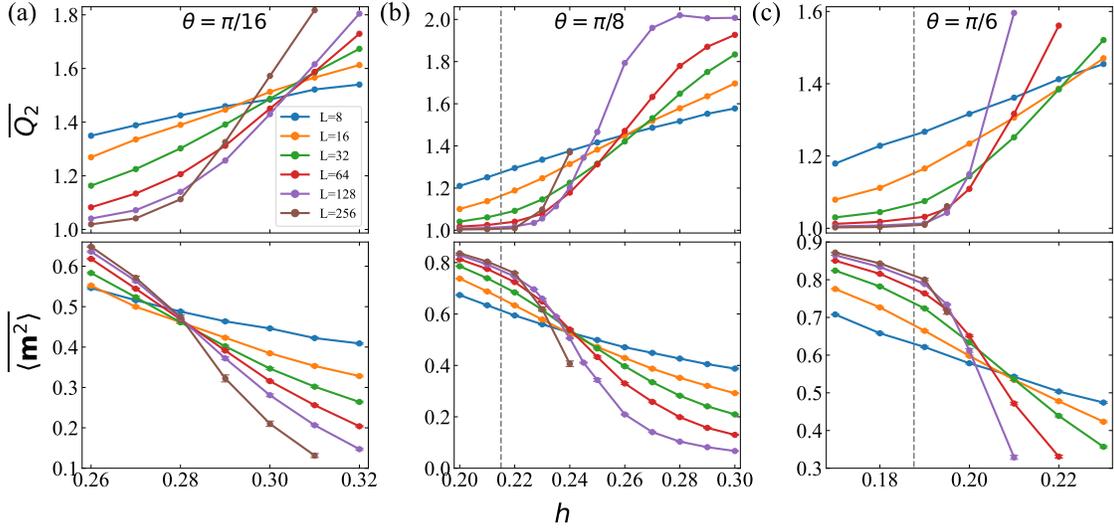

FIG. 5: $\overline{Q_2}$ (upper panels) and $\overline{\langle \mathbf{m}^2 \rangle}$ (lower panels) as function of $h$ for (a) $\theta = \pi/16$ (b) $\theta = \pi/8$ and (c) $\theta = \pi/6$. Dashed lines mark the value of the corresponding clean critical point $h_c^w$. Other parameters: $J_s = 1, J_w = 1/4, p = 0.2$.

calculating $k_{ave}$, see for example Fig. 3(c). On average, $d$ can be estimated as

$$d = \sum_{l_j \geq l_0'} l_j \sim n \sum_{l=l_0'}^{\infty} l f(l-1) = \frac{n}{p}(1-p)^{l_0'}\left(1 + \frac{p}{1-p}l_0'\right). \quad (7)$$

Then the decay of $\overline{A}_r$ is

$$\overline{A}_r \sim \exp\left[-\frac{r}{\xi_0}(1-p)^{l_0'}\left(1 + \frac{p}{1-p}l_0'\right)\right], \quad (8)$$

where we have used the following expression for the average distance

$$r = \sum_{l_j=1}^{n} l_j \sim n\bar{l} = n\sum_{l=1}^{\infty} l f(l-1) = \frac{n}{p}. \quad (9)$$

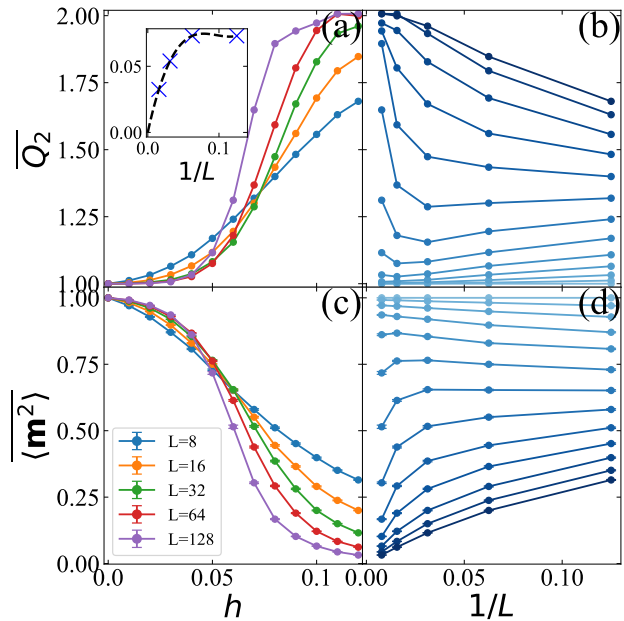

FIG. 6: (a) $\overline{Q_2}$ and (c) $\overline{\langle \mathbf{m}^2 \rangle}$ as function of $h$ at $\theta = \pi/8$ for different $L$. Inset shows the crossing points extracted from $\overline{Q_2}$ curves. A polynomial fit (dashed line) indicates the transition point lies at $h_c = 0$. (b)(d) The same data plotted as a function of $1/L$ for different $h$. Deeper color indicates larger $h$. Parameter used: $J_s = 1$.

So the correlation length for disordered system should be defined as

$$\frac{\xi}{\xi_0} \simeq (1-p)^{-l'_0} \left( 1 + \frac{p}{1-p} l'_0 \right)^{-1}$$

$$\sim (1-p)^{-a\xi_0} \left( 1 + \frac{p}{1-p} a\xi_0 \right)^{-1}. \tag{10}$$

with $a$ some prefactor. Alternatively

$$\ln \frac{\xi}{\xi_0} \sim -a\xi_0 \ln(1-p) - \ln \left( 1 + \frac{p}{1-p} a\xi_0 \right). \tag{11}$$

This expression indicates when approaching the transition from the disordered side $\xi/\xi_0$ shows an essential singularity (the first term) with a logarithmic correction $\propto \ln \xi_0$ (the second term). Neglecting the logarithmic correction

$$\ln \frac{\xi}{\xi_0} \sim -c\xi_0 \ln(1-p) \sim \delta^{-\nu}. \tag{12}$$

Note that if $p = 0$ we recover the clean case $\xi = \xi_0$.

## III. THE C-IC TRANSITIONS AND LOCALIZED DOMAIN WALLS

Here we review the role of domain walls in the low-energy effective theory of the $\mathbb{Z}_N$ CCM and subsequently apply it to discuss the consequence of disorder. We do so by deriving an effective Hamlitonina valid in the vicinity of $\theta = \pi/N$, where we have a small expansion paramter $\delta = \pi/N - \theta$. Our analysis closely follows Refs. [4–6]. In the vicinity of $\pi/N$, CC domain walls ($|10\rangle, |21\rangle, \cdots, |0, N-1\rangle$) possess smaller energy $\epsilon_{cc} = -4J_j \sin \delta \sin(\pi/N) \sim J_j \delta$ compared to all other states. Restricting to the low energy subspace, CC domain walls behave like fermions since they are impenetrable [4, 5]. Up to $O((h/J_j)^2)$ we arrive at the following low energy effective Hamiltonian for spinless fermions (details are given in the last part of this section)

$$H = -\sum_j \mu_j d_j^\dagger d_j - \sum_j h \left( d_{j+1}^\dagger d_j + h.c. \right) - \sum_j U_j n_j n_{j+1} + \sum_j \Delta_j \left( d_{j-1}^\dagger n_j d_{j+1} + h.c. \right), \tag{13}$$

where $d_j$ ($d_j^\dagger$) annihilates (creates) a domain wall at site $j$, $\mu_j = -\epsilon_{cc}$ mimic a random potential and $U_j, \Delta_j$ are the strength of nearest-neighbour interaction and correlated hopping respectively. In the pure case, $\mu_j = \mu$ and $U = 2\Delta$, the interaction and correlated hopping terms can be combined into a four-fermion coupling term, which by power counting is irrelevant to the $\mu = -2h$ critical point [6]. For $\mu < -2h$, the ground state of Eq. (13) is the fermionic vaccum, corresponding to the magnetic ordered phase devoid of domain walls, while for $\mu > -2h$ a Luttinger liquid phase emerges with a finite amount of excess CC domain walls. Therefore across the transition both the magnetization and domain wall density behave like an order parameter of a continuous phase transition. We would like to mention that away from $\theta = \pi/N$, the above qualitative picture remains the same but the critical properties of the transitions are likely to be modified greatly [2]. In the random case, the first two terms of Eq. (13) suggest that excess domain walls are Anderson localized in rare regions. The nearest-neighbour interaction and correlated hopping terms might be irrelevant, since it is known that Anderson localization is (at least perturbatively) stable in the presence weak interactions in 1D [7, 8].

Now we give detailed derivation of the effective Hamiltonian. As discussed above, in the vicinity of $\theta = \pi/N$ and for small $h \ll J_j$, we consider the low-energy subspace containing CC domain walls. In the $\sigma$ representation, we can define the domain wall vacuum as $|0\rangle = |\cdots aaaa \cdots\rangle$ with $a = 0, \ldots, N-1$. We can further introduce the CC domain wall creation operators $d_j^\dagger$, which act as $d_j^\dagger|0\rangle = |\cdots aa, a-1, a-1 \cdots\rangle$. Most importantly, fermionic statistics $\{d_i, d_j^\dagger\} = \delta_{ij}$ are needed to keep CC domain walls from crossing so as to stay in the low-energy subspace. The energy of a CC domain wall measured with respect to the vacuum is

$$\epsilon_j^{cc} = 4J_j \sin \frac{\pi}{N} \sin \delta, \tag{14}$$

with $\delta = \pi/N - \theta$. With the above knowledge, it is easy to see the following mapping

$$-\sum_j J_j \left( \sigma_j^\dagger \sigma_{j+1} e^{i\theta} + \sigma_j \sigma_{j+1}^\dagger e^{-i\theta} \right) \rightarrow H_0 = -\sum_j \mu_j d_j^\dagger d_j, \tag{15}$$

where $\mu_j = -\epsilon_j^{cc}$. Now consider first-order process, the main contribution will be the CC domain wall hopping, say $|\cdots aa, a-1, a-1 \cdots\rangle \rightarrow |\cdots a, a-1, a-1, a-1 \cdots\rangle$, which leads to the Hamiltonian

$$H_1 = -h \sum_j \left( d_{j+1}^\dagger d_j + d_j^\dagger d_{j+1} \right), \tag{16}$$

Two main second-order processes are the nearest-neighbour interaction

$$|\cdots aa, a-1, a-2, a-2 \cdots\rangle \rightarrow |\cdots aa \binom{a-2}{a} a-2, a-2 \cdots\rangle \rightarrow |\cdots aa, a-1, a-2, a-2 \cdots\rangle. \tag{17}$$

and the correlated hopping

$$|\cdots aa, a-1, a-2, a-2 \cdots\rangle \rightarrow |\cdots aa, a-2, a-2, a-2 \cdots\rangle \rightarrow |\cdots aa-1, a-2, a-2, a-2 \cdots\rangle. \tag{18}$$

These processes result in the Hamiltonian

$$H_2 = -\sum_j U_j n_j n_{j+1} + \sum_j \Delta_j \left( d_{j-1}^\dagger n_j d_{j+1} + h.c. \right) \tag{19}$$

where $n_j = d_j^\dagger d_j$ is the density operator and the coupling strengths read

$$U_j = \frac{h^2}{\epsilon_j^h - \epsilon_j^{cc} - \epsilon_{j+1}^{cc}} + \frac{h^2}{\epsilon_{j+1}^h - \epsilon_{j+1}^{cc} - \epsilon_j^{cc}}, \tag{20}$$

and

$$\Delta_j = \frac{h^2}{2} \left( \frac{1}{\epsilon_j^h - \epsilon_j^{cc} - \epsilon_{j+1}^{cc}} + \frac{h^2}{\epsilon_j^h - \epsilon_j^{cc} - \epsilon_{j-1}^{cc}} \right), \tag{21}$$

respectively, with

$$\epsilon_j^h = -2J_j \left( \cos \left( \frac{3\pi}{N} + \delta \right) - \cos \left( \frac{\pi}{N} - \delta \right) \right), \tag{22}$$

being the energy of the state $|\cdots aa, a-2, a-2, \cdots\rangle$. Combining all terms, we arrive at the effective Hamiltonian in Eq. (13).

## IV. DETAILED DERIVATION OF THE DECIMATION RULES OF SDRG

In this section we present detailed derivation of the decimation rules of SDRG. During the renormalization, if the largest coupling is a field, say $h_1$, we split the Hamiltonian concerning spin 1 into the unperturbed part

$$H_0 = -h_1(\tau_1^\dagger e^{i\phi_1} + \tau_1 e^{-i\phi_1}),$$
(23)

and the perturbation

$$H' = -\left(J_0\sigma_0^\dagger\sigma_1 e^{i\theta_0} + J_1\sigma_1^\dagger\sigma_2 e^{i\theta_1}\right) + h.c.$$
(24)

Within second-order perturbation theory, we get the effective Hamiltonian

$$
\begin{aligned}
\langle 0|H_{eff}|0\rangle &= -2h_1\cos\phi_1 + \sum_{\alpha\neq 0}\frac{\langle 0|H'|\alpha\rangle\langle\alpha|H'|0\rangle}{E_\alpha - E_0} \\
&= -2h_1\cos\phi_1 + \frac{(J_0\sigma_0^\dagger e^{i\theta_0} + J_1\sigma_2^\dagger e^{-i\theta_1})(J_0\sigma_0 e^{-i\theta_0} + J_1\sigma_2 e^{i\theta_1})}{-2h_1\cos\phi_1 + 2h_1\cos(\phi_1 + 2\pi/N)} \\
&\quad + \frac{(J_0\sigma_0 e^{-i\theta_0} + J_1\sigma_2 e^{i\theta_1})(J_0\sigma_0^\dagger e^{i\theta_0} + J_1\sigma_2^\dagger e^{-i\theta_1})}{-2h_1\cos\phi_1 + 2h_1\cos(\phi_1 - 2\pi/N)} \\
&= -2h_1\cos\phi_1 - \frac{J_0^2 + J_1^2}{h_1}\kappa_{\phi_1} - \frac{J_0 J_1}{h_1}\kappa_{\phi_1}\left(\sigma_0^\dagger\sigma_2 e^{i(\theta_0 + \theta_1)} + h.c.\right),
\end{aligned}
$$
(25)

where $|\alpha\rangle$ are the eigenstates of $\tau$ and

$$\kappa_{\phi_1} = \frac{1}{2}\left(\frac{1}{\cos\phi_1 - \cos(\phi_1 - 2\pi/N)} + \frac{1}{\cos\phi_1 - \cos(\phi_1 + 2\pi/N)}\right).$$
(26)

On the other hand, if the largest coupling is a link, say $J_1$, the unperturbed Hamiltonian is

$$H_0 = -J_1\left(\sigma_1^\dagger\sigma_2 e^{i\theta_1} + \sigma_1\sigma_2^\dagger e^{i\theta_1}\right),$$
(27)

and the perturbation term is

$$H' = -\left(h_1\tau_j^\dagger e^{i\phi_1} + h_2\tau_2^\dagger e^{i\phi_2}\right) - \left(J_0\sigma_0^\dagger\sigma_1 e^{i\theta_0} + J_2\sigma_2^\dagger\sigma_3 e^{i\theta_2}\right) + h.c.$$
(28)

Applying the perturbation theory, we arrive at

$$
\begin{aligned}
\langle\beta\beta|H_{eff}|\alpha\alpha\rangle &= -2J_1\cos\theta_1 + \langle\beta\beta|H'|\alpha\alpha\rangle + \sum_{\gamma\neq\gamma'}\frac{\langle\beta\beta|H'|\gamma\gamma'\rangle\langle\gamma\gamma'|H'|\alpha\alpha\rangle}{2}\left(\frac{1}{E_{\beta\beta} - E_{\gamma\gamma'}} + \frac{1}{E_{\alpha\alpha} - E_{\gamma\gamma'}}\right) \\
&= -2J_1\cos\theta_1 - \delta_{\alpha\beta}\left(J_0\sigma_0^\dagger\omega^\alpha e^{i\theta_0} + J_2\sigma_3\omega^{-\alpha}e^{i\theta_2} + h.c.\right) - \frac{h_1^2 + h_2^2}{J_1}\kappa_{\theta_1} \\
&\quad - \frac{h_1 h_2}{J_1}\kappa_{\theta_1}\left(\delta_{\beta,\alpha+1}e^{-i(\phi_1 + \phi_2)} + \delta_{\beta,\alpha-1}e^{i(\phi_1 + \phi_2)}\right),
\end{aligned}
$$
(29)

where $|\alpha\rangle$ are the eigenstates of $\sigma$ and

$$\kappa_{\theta_1} = \frac{1}{2}\left(\frac{1}{\cos\theta_1 - \cos(\theta_1 - 2\pi/N)} + \frac{1}{\cos\theta_1 - \cos(\theta_1 + 2\pi/N)}\right).$$
(30)

Replacing $0 \to j-1, 1 \to j, 2 \to j+1$, we arrive at the general decimation rules in the main text.

---



\end{document}